\documentclass{PoS}
\usepackage{amsmath}
\usepackage{graphicx}
\usepackage{esint}
\usepackage[square,comma,numbers,sort&compress]{natbib}
\def\tb{}

\title{Soft Matrix Elements in Non-local Chiral Quark Model}

\ShortTitle{Soft Matrix Elements in Non-local Chiral Quark Model}

\author{Piotr Kotko\thanks{Work done in collaboration with M. Praszalowicz.}
\footnote{The poster is available at http://th-www.if.uj.edu.pl/\textasciitilde kotko/files/posters/posterEPS.pdf}\\
        Jagiellonian University\\
        E-mail: \email{kotko@th.if.uj.edu.pl}}

\abstract{Using non-local chiral quark model and currents satisfying Ward-Takahashi identities
we analyze Distribution Amplitudes (DA) of photon and pion-to-photon Transition Distribution Amplitudes (TDA) in
the low energy regime.
Photon DA's are calculated analytically up to twist-4 and reveal several interesting features of
photon structure.
TDA's calculated in the present model satisfy polynomiality condition. Normalization of
vector TDA is fixed by the axial anomaly.
We also compute relevant form factors and compare them with existing data.
Axial form factor turns out to
be much lower then the vector one, what indeed is seen in the experimental data.}

\FullConference{European Physical Society Europhysics Conference on High Energy Physics,
EPS-HEP 2009,\\
		 July 16 - 22 2009\\
		 Krakow, Poland}

\begin{document}

\section{Introduction}
\vspace{-0.2cm}
Exclusive processes serve as
important testing ground both for perturbative and non-perturbative
QCD. Due to factorization theorems
pertinent amplitudes factorize into hard parts calculable in
perturbative QCD
and non-perturbative objects parameterizing matrix elements of 
non-local operators on the light cone. The simplest objects of this
kind are Distribution Amplitudes (DA) which correspond to vacuum-to-hadron
matrix elements (ME) of quark or gluon non-local operators. They
appear \textit{e.g.} in QCD expressions for hadronic form factors
\citep{Radyushkin,Brodsky}

More general class of such objects is called Transition DA's and was introduced in \cite{Pire}. They parametrize  ME of bilocal operators {\em non-diagonal in momenta and in physical states}. Since they involve momentum transfer, they provide also information on the transverse structure.

In the present note we employ low energy effective
model to study DA's and TDA's. This is, however, non-trivial since such models in general
do not inherit all symmetries of QCD. This can cause problems because
all low energy amplitudes have to possess several properties originating
from QCD, \textit{e.g.} they have to obey Lorentz invariance and Ward
identities, or reproduce anomalies.

If we limit hadronic states to the pions
only, we can analyze soft matrix elements using semi-bosonized Nambu-Jona-Lasinio
model, which incorporates spontaneous chiral symmetry breaking
resulting in momentum
dependent constituent quark mass $M\left( k\right)$. For present calculations we choose the form of
$M\left( k\right)$ proposed in \cite{Rostw}.
Due to this
momentum dependence Ward-Takahashi identities are violated. In order to fix this deficiency the standard vertices
$\gamma^{\mu}$ and $\gamma^{\mu}\gamma_{5}$ have to be replaced by the non-local ones, which however cannot be determined uniquely \cite{Holdom,Bowler}. In the present calculations we use the "minimal" form of the non-local vertices given for example in \cite{Holdom}.

\vspace{-0.2cm}
\section{Photon Distribution Amplitudes}
\vspace{-0.2cm}

One of the simplest applications of the
non-local NJL model are DA's. Although
they are usually defined for hadrons, one can also consider DA's of
the photon, due to its hadronic part. They appear for example in description
of the radiative meson decays \citep{Ball,Bronfoton,Polyakov,foton}.

Using light cone coordinates defined
by two null vectors $\tb{n=\left(1,0,0,-1\right)}$ and $\tb{\tilde{n}=\left(1,0,0,1\right)}$,
the general definition of photon DA's can be written as
\begin{equation}
\int\frac{d\lambda}{2\pi}\, e^{i(2u-1)\lambda P^{+}}\left\langle 0\left\vert \overline{\psi}\left(\lambda n\right)\mathcal{O}\psi\left(-\lambda n\right)\right\vert \gamma\left(P\right)\right\rangle\sim F_{\mathcal{O}}\left(P^{2}\right)
\sum_{\tau}\,\mathcal{O}_{\tau}\,\,\phi_{\mathcal{O}}^{(\tau)}\left(u,P^{2}\right) ,
\end{equation}
where $\tb{P^{\mu}}$ is the momentum of the photon, $\tb{P^{+}=P\cdot n}$ and $\tb u$ is longitudinal momentum
fraction of a quark in the photon.
Dirac matrices $\tb{\mathcal{O}=\left\{ \sigma^{\mu\nu},\gamma^{\mu},\gamma^{\mu}\gamma_{5}\right\} }$
correspond to different tensor nature of bilocal operators. $\mathcal{O}_{\tau}$
denotes appropriate tensor structure standing in front of DA $\tb{\phi_{\mathcal{O}}}$
of given dynamical twist $\tau$.
$\tb{F_{\mathcal{O}}\left(P^{2}\right)}$
is the relevant momentum dependent decay constant.

Using the model described in the Introduction
 we calculated {\em analytically} (up to numerical solution of certain algebraic equation) 
 photon DA's of twist 2,3 and 4 \cite{foton}. Leading twist tensor DA turns out to be almost flat and does not vanish at the endpoints. 
 In the vector and axial channels the relevant matrix elements
 possess  additional
perturbative non-hadronic component which has to be subtracted. Higher twist DA's are rather strongly model dependent and some of them possess delta-type singularities at the end-points.
\vspace{-0.2cm}
\section{Transition Distribution Amplitudes}
\vspace{-0.2cm}

Transition Distribution Amplitudes (TDA) were originally introduced
in order to describe processes like hadron-antihadron annihilation
into 
photons $\tb{\bar{H}H\rightarrow\gamma^{*}\gamma}$
or backward virtual Compton scattering \cite{Pire}. Amplitudes for these
reactions factorize similarly as for $\tb{\bar{H}H\rightarrow\gamma^{*}}$,
with the restriction that usual DA's for hadrons
should be replaced by TDA's.

In practice we deal with two kinds
of twist-2 TDA's: vector (VTDA) and axial
 (ATDA) \citep{Pire,Tiburzi,BronTDA,Noguera,TDA_1,TDA_2}. They can be schematically defined as
\vspace{-0.1cm}
\begin{equation}
\int\frac{d\lambda}{2\pi}e^{i\lambda Xp^{+}}\left\langle \gamma\left(P_{2},\varepsilon\right)\left|\overline{d}\left(-\lambda n\right)\mathcal{O}\, u\left(\lambda n\right)\right|\pi^{+}\left(P_{1}\right)\right\rangle \sim \mathcal{O}_{\mathrm{twist-2}}\,\, D\left(X,\xi,t\right),
\end{equation}
\vspace{-0.1cm}
where $t=q^2=\left( P_2 - P_1\right)^{2}$,
$p=\frac{1}{2}\left(P_{1}+P_{2}\right)$ and $\xi=-q^{+}/2p^{+}$ is so called skewedness. $D$ should be replaced by $V$ or $A$ in vector or axial channels respectively ({\em i.e.} if $\mathcal{O}=\gamma^{\mu}$ or $\mathcal{O}=\gamma^{\mu}\gamma_5$).

\begin{figure}
\begin{tabular}{ll}
a)\vspace{-0.8cm} & b)\tabularnewline
\includegraphics[height=4.cm]{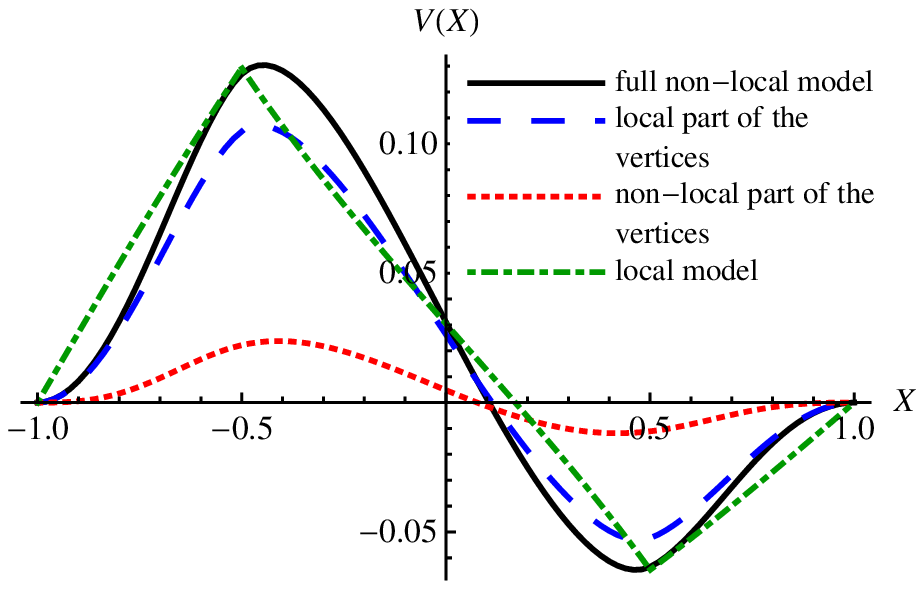} & \qquad
\includegraphics[height=4.cm]{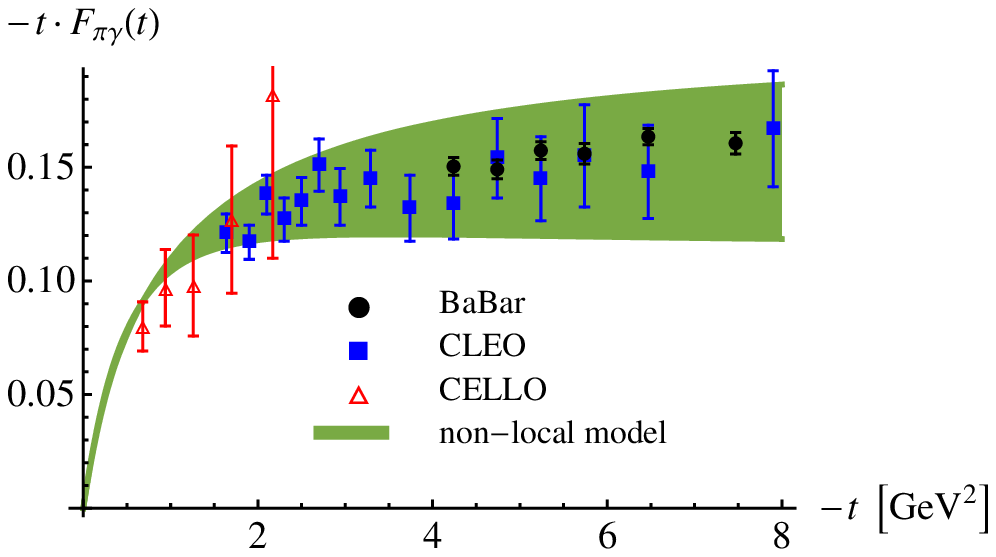}
\tabularnewline
\end{tabular}
\vspace{-0.4cm}
\caption{a) Model prediction  for  VTDA at $t=-0.1 \mathrm{GeV}^2$  (solid) is the
sum of the pieces coming from the local part of the vertex (dashed) and the non-local one (dotted).
b) Pion-photon transition form factor and the data. Shaded area corresponds to different choices of model parameters.}%
\vspace{-0.1cm}
\label{figure}%
\end{figure}

TDA's should satisfy  polynomiality condition following
from Lorentz invariance. This constraint turns out to be satisfied in our model.
Normalization of VTDA is fixed once for all by axial anomaly and equals
$\int dX\, V\left(X,\xi,t=0\right)=1/2\pi^{2}$,
while the normalization of ATDA remains unconstrained.
Using modified vertices we are able to recover correct normalization
for VTDA in the non-local model \cite{TDA_2}. As an example we plot
VTDA in Fig. \ref{figure}a.

Zeroth moments of VTDA's and ATDA's
are related to vector $F_V(t)$ and axial $F_A(t)$ form factors. Model independent prediction
for $\tb{F_{V}\left(0\right)}$ due to the axial anomaly reads:
$F_{V}\left(0\right)\approx0.027$
which overshoots the experimental value given by PDG
$F_{V}^{\mathrm{exp}}\left(0\right)=0.017\pm0.008$.
In the case of axial form factor local models (\textit{i.e.} with $M\left(k\right)\equiv \mathrm{const}$) predict that $\tb{F_{A}\left(0\right)=F_{V}\left(0\right)}$.
However experimentally $\tb{F_{A}\left(0\right)}$ is
much smaller. Values given by PDG read:
$F_{A}^{\mathrm{exp}}\left(0\right)=0.0115\pm0.0005$ and $\left(F_{A}\left(0\right)/F_{V}\left(0\right)\right)_{\mathrm{exp}}=0.7_{-0.2}^{+0.6}$ for the ratio.
In our model we get indeed lower values
for $\tb{F_{A}\left(0\right)}$. We obtain $F_A(0)=0.0152\div 0.0217$ and
$F_{A}\left(0\right)/F_{V}\left(0\right)=0.56\div 0.80$ depending on model parameters.

There is another very important quantity directly related to $F_{V}$,
namely pion to photon transition form factor $\tb{F_{\pi\gamma}\left(t\right)}$
which controls $\tb{\pi^{0}\rightarrow\gamma^{*}\gamma}$ reactions.
It was measured by CELLO, CLEO and recently by BaBar \citep{Cello,Cleo,BaBar}.
We compare our predictions with data  in Fig. \ref{figure}b  in low momentum
transfer regime where our model should apply.

\end{document}